\documentclass{elsart}
\usepackage{epsf}

\def\figuresize{12.5cm}

\usepackage{amssymb}

\begin{document}

\begin{frontmatter}



\title{The Hartree-Fock Based Diagonalization\\
       - an Efficient Algorithm for the Treatment\\
       of Interacting Electrons in Disordered Solids}


\author[c,b]{Michael Schreiber}
and
\author[c,o]{Thomas Vojta}

\address[c]{Institut f{\"u}r Physik, Technische Universit{\"a}t
            Chemnitz, D-09107 Chemnitz, Germany}
\address[b]{School of Engineering and Science, International University
            Bremen, Germany}
\address[o]{Theoretical Physics, University of Oxford, 1 Keble Road,
            Oxford OX1 3NP, UK}

\begin{abstract}
The Hartree-Fock based diagonalization is a computational method for
the investigation of the low-energy properties of correlated electrons
in disordered solids. The method is related to the quantum-chemical
configuration interaction approach. It consists in diagonalizing the
Hamiltonian in a reduced Hilbert space built of  the low-energy
states of the corresponding disordered Hartree-Fock Hamiltonian.
The properties of the method are discussed for the example of the
quantum Coulomb glass, a lattice model of  electrons in a random potential
interacting via long-range Coulomb interaction. Particular attention is
paid to the accuracy of the results as a function of the dimension
of the reduced Hilbert space. It is argued that disorder actually
helps the approximation.
\end{abstract}

\end{frontmatter}

\section{Introduction}
In recent years there has been a strong interest in the electronic properties
of correlated electrons in the presence of disorder, e.g., in connection
with the fractional quantum Hall effect, heavy-fermion systems or the
apparent metal-insulator transition of the two-dimensional electron gas in
Si-MOSFETs. If both disorder and interactions are strong,
the usual analytic theories often do not work well.
For instance, a complete analytic description of the localized, insulating phase
of disordered interacting electrons has not been achieved. Therefore,
computational methods are particularly important in this area.

However, the numerical simulation of disordered many-particle systems is one
of the most complicated problems in computational condensed matter physics.
First, the size of the many-particle Hilbert space to be considered grows
exponentially with the system size. Second, in the presence of disorder
many samples with different disorder configurations have to be simulated
in order to obtain averages or typical values of physical quantities.
Since many interesting quantities like the conductance are non-self-averaging,
often the entire distribution function of an observable is desired,
requiring at least of the order of 10$^3$ samples.
Moreover, the basic interaction between the electrons, the Coulomb interaction,
is long-ranged. While in the metallic phase it is often possible to use an
effective model with screening already built in, the long-range Coulomb interaction
has to be retained for a correct description of the insulating phase
when screening breaks down.
This drastically increases the numerical effort required.

One way to overcome the problem of the exponentially (in the number of particles)
large Hilbert space is to use mean-field ideas to reduce the system
to a self-consistent effective single-particle problem. The simplest way to
do this are the Hartree and Hartree-Fock approximations
(see \cite{hf,epper_hf} for applications of the Hartree-Fock approximation
to lattice models of disordered interacting electrons).
In the more sophisticated form of the density functional theory this effective
single-particle approach has proven to be successful in predicting the
electronic structure of a wide variety of materials.
In general, the effective single-particle methods permit the simulation of
rather large systems ($>10^3$ sites) but the approximations involved are
uncontrolled and can usually not be improved systematically.
Effective single-particle methods usually work well on energy scales of eV.
At low temperatures and energies of the order of meV electronic correlations
become more important and new physical phenomena beyond the scope of effective
single-particle theories emerge. Here methods are desired which  give
numerically exact results or which can be taken, at least in principle,
to arbitrary accuracy. However, most of these methods are severely
restricted when simulating disordered interacting electrons.

Exact diagonalization requires the diagonalization of a matrix whose
dimension is that of the full many-particle Hilbert space. Therefore,
it works only for very small systems (e.g., with up to about $4 \times 4$ lattice
sites in two dimensions) \cite{exact,epper_exact}.
For one-dimensional systems the density-matrix renormalization
group (DMRG) method \cite{dmrg} is a very efficient tool to obtain the
low-energy properties. It is, however, less effective in higher dimensions.
Moreover, the long-range Coulomb interaction strongly complicates the original
real-space based DMRG. Later, momentum-space based versions of the DMRG
method have also been tested \cite{dmrgp}.
However, it turned out that their accuracy
is generically less than that of the the real-space DMRG under favorable
conditions \cite{dmrgt}.
Quantum Monte-Carlo \cite{qmc} methods are another means of simulating
disordered many-particle systems. They are very effective for bosons at
finite temperatures. Very low temperatures are, however, hard to reach.
Moreover, simulations of fermions suffer from the notorious sign problem,
although this turned out to be less severe in the presence of disorder.
Recently, the sign-problem could be further reduced using a
multilevel blocking Monte-Carlo method \cite{blockqmc}.

In this paper we present an alternative method for simulating disordered
interacting electrons, the Hartree-Fock based diagonalization (HFD). It is
related to the configuration interaction method \cite{CI}. This is one of the early
methods for treating interacting electrons which is particularly popular
in quantum chemistry. We will show that the Hartree-Fock based diagonalization
is a very suitable method for strongly disordered interacting electrons.
In fact, the algorithm benefits from the presence of disorder. The remainder of
this paper is organized as follows. In Sec.\ \ref{sec:HFD} we present the
method and discuss its properties. In Sec.\ \ref{sec:QCG} we show the results of
various test calculations for the quantum Coulomb glass Hamiltonian.
Section \ref{sec:Results} summarizes a few important physical results obtained
using the HFD method. We conclude in Sec.\ \ref{sec:CON}.

\section{Hartree-Fock based diagonalization}
\label{sec:HFD}
The Hartree-Fock based diagonalization (HFD) is similar to the configuration
interaction method \cite{CI} used in quantum chemistry, but
adapted for disordered lattice Hamiltonians. The main idea of the HFD is
to diagonalize the Hamiltonian in a small subspace of the Hilbert space spanned
by the low-energy eigenstates of the corresponding Hartree-Fock approximation.
A schematic of the Hartree-Fock based
diagonalization method is shown in Fig.\ \ref{fig:HFD}.
\begin{figure}[b]
\begin{center}
\fboxrule1pt
\fboxsep5mm
\fbox{\parbox[c]{140mm}{
\begin{tabbing}
\= do for \=each disorder configuration \hfill\\
\>   \>  solve  HF approximation\\
\>   \>  construct many-particle HF states\\
\>   \>  find lowest-in-energy HF states \\
\>   \>  transform Hamiltonian to basis of low-energy HF states\\
\>   \>  diagonalize Hamiltonian\\
\>   \>  transform observables to HF basis and calculate their values\\
\> enddo
\end{tabbing}
}}
\end{center}
\caption{Structure of the Hartree-Fock (HF) based diagonalization method.}
\label{fig:HFD}
\end{figure}
The first step of the HFD method consists in solving the Hartree-Fock
approximation of the Hamiltonian. Due to the presence of disorder this is
a non-trivial single-particle problem which can only be solved numerically.
In the second step many-particle Slater states are constructed from the
single-particle states of the Hartree-Fock approximation. An elaborate Monte-Carlo
algorithm is then used to find the many-particle states with
the lowest energy expectation values. In the third step the Hamiltonian
is transformed into the basis formed by these states and diagonalized.
Finally, the operators of the desired observables are also transformed
into the new basis, and their values are calculated.

Since the Hartree-Fock states are comparatively close in character to the
exact eigenstates in the entire parameter space the Hartree-Fock based
diagonalization works well for all parameters while related methods
based on non-interacting or classical eigenstates \cite{efros95,talamantes96}
instead of Hartree-Fock states are restricted to small parameter regions.
The presence of strong disorder actually helps the method because it breaks
possible symmetries in the Hamiltonian and lifts the related degeneracies
of the eigenenergies which usually pose severe problems for Hartree-Fock
calculations.

In the following we will illustrate the application of the HFD method for the
example of the quantum Coulomb glass, a model of interacting electrons in a
random potential. The model is defined on a regular hypercubic lattice with
$\mathcal{N}=L^d$
($d$ is the spatial dimensionality)
sites occupied by $N=K \mathcal{N}$ electrons ($0\!<\!K\!<\!1$). To ensure charge neutrality
each lattice site carries a compensating positive charge of  $Ke$. The Hamiltonian
is given by
\begin{equation}
H =  -t  \sum_{\langle ij\rangle} (c_i^\dagger c_j + c_j^\dagger c_i) +
       \sum_i \varphi_i  n_i + \frac{1}{2}\sum_{i\not=j}(n_i-K)(n_j-K)U_{ij}
\label{eq:Hamiltonian}
\end{equation}
where $c_i^\dagger$ and $c_i$ are the electron creation and annihilation operators
at site $i$, respectively,  and $\langle ij \rangle$ denotes all pairs of nearest
neighbor sites.
$t$ is the strength of the hopping term, i.e., it corresponds to the kinetic energy,
and $n_i$ is the occupation number of site $i$.
We parametrize the Coulomb interaction $U_{ij} = e^2/r_{ij}$ by its value $U$
between nearest neighbor sites. For a correct description of the insulating phase the Coulomb
interaction has to be kept long-ranged, since screening breaks down in the insulator.
The random potential values $\varphi_i$ are chosen
independently from a box distribution of width $2 W_0$ and zero mean.
For $U_{ij}=0$ the quantum Coulomb glass becomes identical to the Anderson model of
localization  and for $t=0$ it turns into  the classical Coulomb glass.
We note that (\ref{eq:Hamiltonian}) describes a system of spinless electrons.
However, the inclusion of the electron spin is straightforward (see Sec.\
\ref{sec:Results}; it just doubles the number of degrees of freedom).

We now give a more  detailed description of the HFD method for the quantum
Coulomb glass.
For each disorder configuration the first step consists in numerically
diagonalizing the Hartree-Fock approximation
\begin{eqnarray}
H_{\rm HF} = &-t  & \sum_{\langle ij\rangle}  (c_i^\dagger c_j + c_j^\dagger c_i)
+  \sum_i  \varphi_i  n_i \nonumber \\
 &+ & \sum_{i\not=j}  n_i ~ U_{ij} \langle n_j -K \rangle
- \sum_{i,j} c_i^\dagger c_j ~ U_{ij} \langle c_j^\dagger c_i \rangle,
\label{eq:HF}
\end{eqnarray}
of the Hamiltonian as described in Ref. \cite{epper_hf}. Here
$\langle \ldots \rangle$ represents the expectation value with respect to
the Hartree-Fock ground state which has to be determined self-consistently.
This calculation results in an orthonormal set of single-particle
Hartree-Fock states
$|\psi_\nu\rangle=b_\nu^\dagger|0\rangle$ which defines a unitary transformation
$b_\nu^\dagger=\sum_i S_{\nu i}c_i^\dagger$ .

In the second step of the HFD method we construct many-particle states,  i.e., Slater determinants,
\begin{equation}
  |\{\nu\} \rangle = b_{\nu_1}^\dagger \ldots b_{\nu_N}^\dagger | 0 \rangle~.
\label{eq:MP_states}
\end{equation}
Note that for the two limiting cases mentioned above,
i.e for the Anderson model of localization and
for the classical Coulomb glass, these states are also eigenstates of the full
Hamiltonian (\ref{eq:Hamiltonian}).
We then determine the set of  many-particle states
(\ref{eq:MP_states}) which have the lowest expectation values
$\langle \{\nu\}|H | \{\nu\}\rangle$ of the energy.
This set varies from disorder realization to disorder realization in
a non-trivial way making it difficult to use an efficient deterministic
algorithm for this task. Since the total number
of states is too high for a complete enumeration we employ a Monte-Carlo method.
It is based on the thermal cycling method \cite{cycling} in which the system is
repeatedly heated and cooled. In addition, at the end of each cycle a systematic
local search around the current configuration is performed.
While performing the Monte-Carlo simulation we keep an archive containing
the $B$ states with the lowest energies encountered so far. When no new
low-energy states have been found for a certain number of Monte-Carlo steps
we stop.
The set of low-energy many-particle states  found in this way
spans the sub-space of the Hilbert space relevant for the low-energy
properties. Its dimension $B$ determines the accuracy of the results.
(If we choose $B$ equal to the full Hilbert space dimension the results
will be exact since all we have done is a unitary transformation).

The third HFD step consists in transforming the Hamiltonian from the original site representation
to the Hartree-Fock representation and calculating the matrix elements
$\langle \{\nu\}|H| \{\mu\}\rangle$. The resulting Hamiltonian matrix $H_{\{\nu\}\{\mu\}}$
of size $B\times B$ is then diagonalized using standard library routines.
Note that $H_{\{\nu\}\{\mu\}}$ is usually {\em not} very sparse: if  $| \{\nu\}\rangle$ and
$| \{\mu\}\rangle$ differ in the occupation of at most 4 single-particle states, the
matrix element  is non-zero. Moreover, number and position of the non-zero
matrix elements differ between different disorder configurations. Thus, specialized
codes for sparse matrices will not increase the performance significantly.
The diagonalization gives the eigenenergies and the eigenstates in the
Hartree-Fock basis $| \{\nu\}\rangle$. In order to calculate physical
observables
like occupation numbers or transport properties we transform their operators
to the Hartree-Fock representation. This is usually faster than transforming
the eigenstates states back to site or momentum representation.

\section{Test calculations for the quantum Coulomb glass model}
\label{sec:QCG}

\begin{figure}
  \epsfxsize=\figuresize
  \centerline{\epsffile{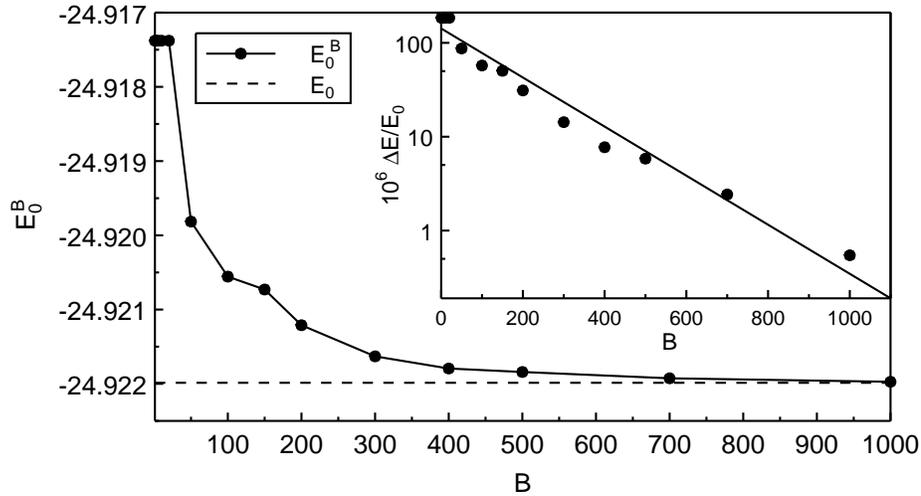}}
  \caption{Dependence of the ground state energy $E_0^B$ and its
              relative error $(E_0^B-E_0)/E_0$
              on the size $B$ of the basis used for a system of 8 electrons
              on $4 \times 4$ sites, $W_0=1,~t=0.1,~U=1$. The solid line in the inset
              is a fit to an exponential law. For comparison, the energy of the first excited
              state is $E_1=-24.73$.}
  \label{fig:ground_state_energy}
\end{figure}

In order to test the method and to check the dependence of the results on
the size $B$ of the basis we have carried out extensive simulations for
systems with $4 \times 4$ sites. We have compared the results
to those of exact diagonalizations which are easily possible for spinless
electrons on a $4 \times 4$ lattice.
First we have investigated the dependence of the ground state energy $E_0^B$
on $B$ and compared it with the exact result $E_0$. A typical example of such a
calculation for one particular disorder realization is presented
in Fig. \ref{fig:ground_state_energy}.
As usual the ground state energy is not very sensitive to the accuracy of the
approximation. Already the relative energy error of the Hartree-Fock approximation
($B=1$ in Fig. \ref{fig:ground_state_energy}) is as low as $2 * 10^{-4}$. This is
about 2.5\% of the energy separation between the ground and the first excited
state.
Keeping a basis size of 300 within the HFD method reduces the error by
a factor of 10. Further increasing the basis size to 1000 (which is still
less than 10\% of the total Hilbert space dimension of 12870)
gives a relative  error of less than $10^{-6}$. For the basis sizes
covered in our calculation the
convergence of the ground state energy is approximately exponential
(see inset of Fig. 2).

\begin{figure}
  \epsfxsize=\textwidth
  \centerline{\epsffile{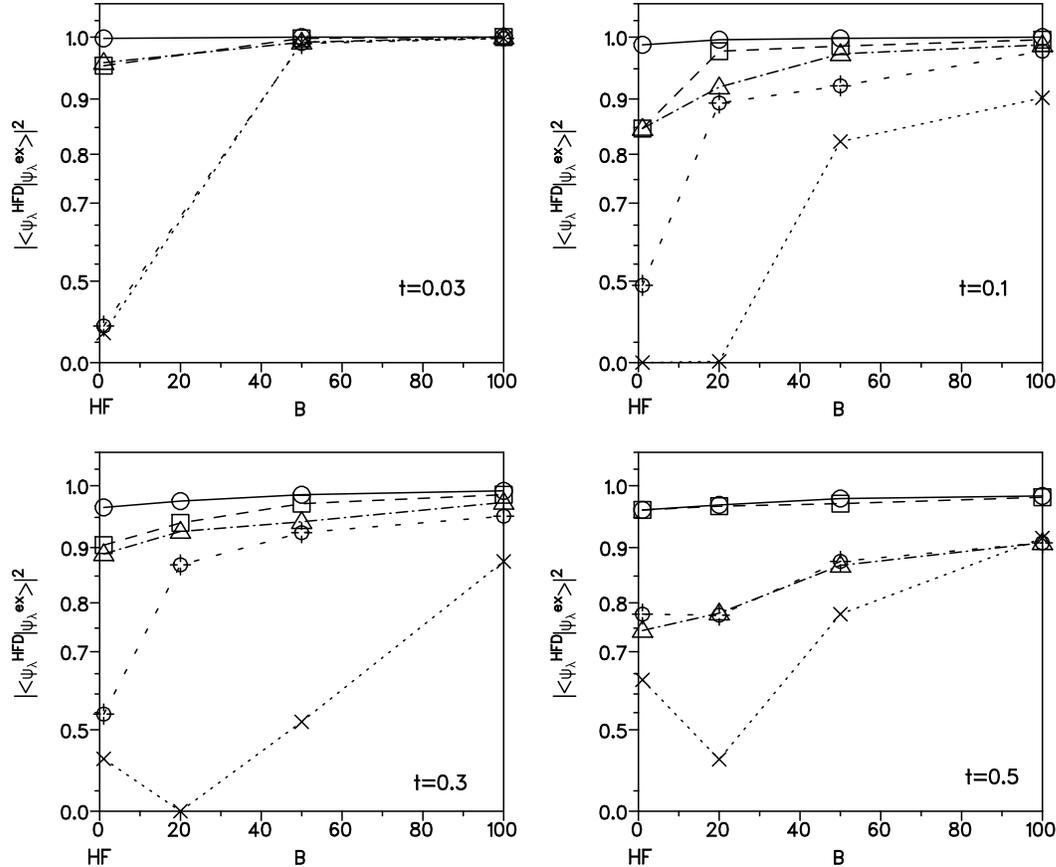}}
  \caption{Overlap between the HFD eigenstates and the corresponding exact
           eigenstates for a system of 8 electrons on $4 \times 4$ sites,
           $W_0=1,~U=1$ and several values of $t$. Ground state: $\circ$,
           1st to 4th excited state: $\square\bigtriangleup\oplus\times$.
           Note that the overlap axis has a quadratic scale.}
  \label{fig:overlap}
\end{figure}

Since the ground state energy is not a very sensible indicator to judge the
quality of the approximate ground state we also studied the overlap between the
approximate and the exact eigenstates. Results for one disorder realization
for several parameter sets are shown in Fig. \ref{fig:overlap}.
In all cases a basis size of $B=100$ is sufficient to achieve an overlap
larger than 0.99 between the approximate and the exact ground states.
The approximation works very well for small $t$ because the system is close
to the classical limit ($t=0$) where the Hartree-Fock approximation
will become exact. It will also work well for $t\gg 1 $ when the system
approaches the non-interacting limit because the Hartree-Fock approximation
is exact in this limit, too. The excited states in general converge slower
than the ground state. For $t=0.3$ and 0.5 a basis size of $B=100$ is not
sufficient to obtain more than the first excited state with an accuracy
comparable to that of the ground state.

\begin{figure}
  \epsfxsize=\figuresize
  \centerline{\epsffile{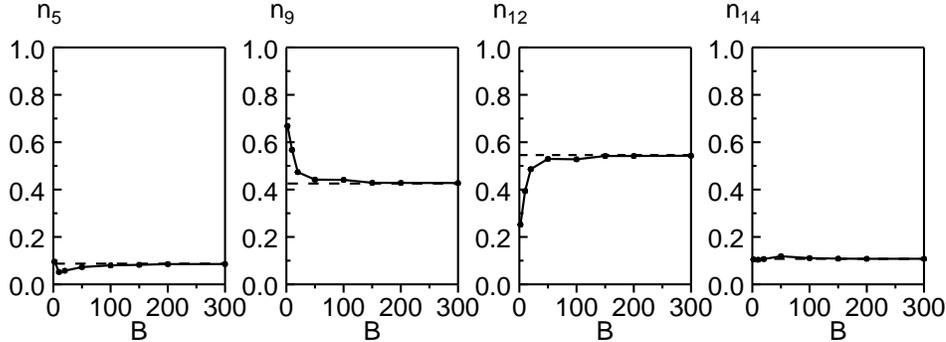}}
  \caption{Ground state occupation numbers $\langle n_i \rangle$ of selected
           sites $i$  vs. basis size $B$  for a system of 8 electrons on
           $4 \times 4$ sites, $W_0=1,~t=0.1,~U=1$.}
  \label{fig:occ}
\end{figure}

In addition to the ground state energy and the overlaps between the approximate
and exact eigenstates we have investigated the convergence properties of several
ground state expectation values. Figure \ref{fig:occ} shows the
convergence of the occupation numbers as a function of basis size $B$
for the same system as in Fig.\ \ref{fig:ground_state_energy}.
While some of the occupation numbers have significant errors within Hartree-Fock
approximation, the HFD method with a basis size of 100 gives all occupation numbers
with a satisfactory accuracy of better than $10^{-2}$.

\begin{figure}
  \epsfxsize=\textwidth
  \centerline{\epsffile{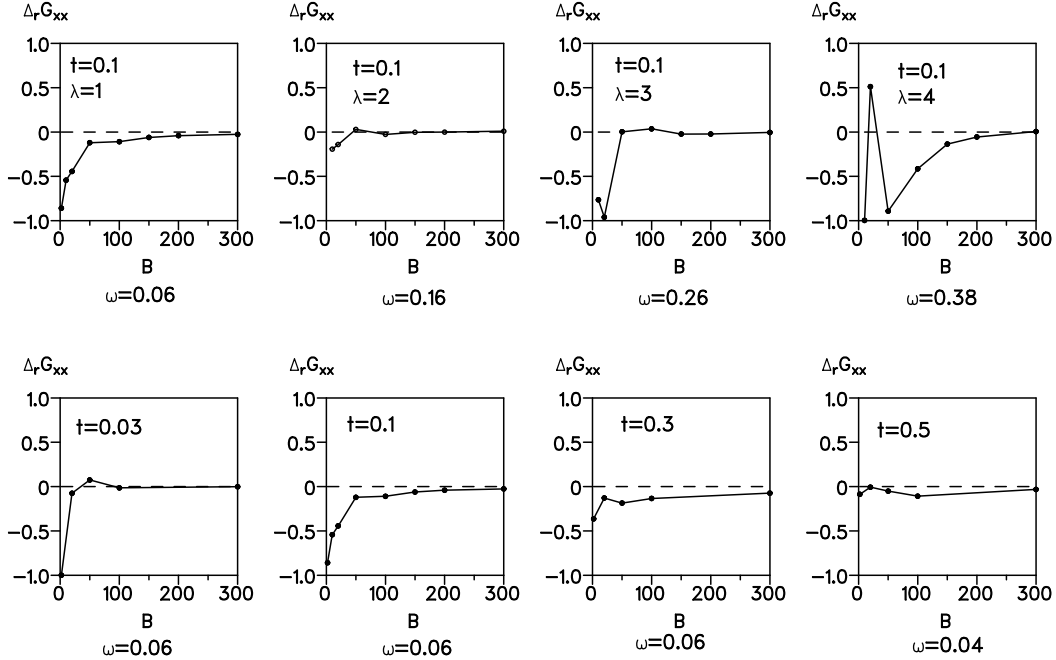}}
  \caption{Relative error of the conductance $G(\omega)$ as a function of
   basis size $B$ for a system of 8 electrons on  $4 \times 4$ sites, $W_0=1,~
   U=1$. Upper row: $t=0.1$, different frequencies ($\lambda$ is the number
   of the contributing excited state), lower row: contribution
   of the first excited state for different $t$.}
  \label{fig:sig_conv}
\end{figure}

As a last example for the comparison between HFD and exact results we want
to discuss the frequency dependent conductance. It is calculated from the
Kubo-Greenwood formula \cite{kubo_greenwood} which relates the conductance
to the current-current correlation function in the ground state.
Using the spectral representation of the correlation function the real
(dissipative) part
of the conductance (in units of the quantum conductance $e^2/h$)
is obtained as
\begin{equation}
 \Re ~ G^{xx}(\omega) = \frac {2 \pi^2}   { \omega}  L^{d-2} \sum_{\alpha} |\langle 0 | j^x|\alpha \rangle |^2
     \delta(\omega+E_0-E_{\alpha})
\label{eq:kubo}
\end{equation}
where $j^x$ is the $x$ component of the current operator and $\alpha$ denotes the eigenstates
of the Hamiltonian.  The finite life time $\tau$ of the eigenstates in a real d.c.\ transport experiment
results in an inhomogeneous broadening $\gamma = 1/\tau$
of the $\delta$ functions in the Kubo-Greenwood formula. Here we have
chosen $\gamma=0.05$ which is of the order of the single-particle level spacing.
While we expect the conductance values to slightly vary with $\gamma$ the
qualitative dependence on disorder or interaction strength will be independent
of the broadening value.
According to eq.\ (\ref{eq:kubo}) the accuracy of the conductance depends not only on
the accuracy of the ground state but also on that of the excited states.
For this reason, the conductance is very sensitive to the quality of the
approximation. Figure \ref{fig:sig_conv} shows the relative error of the
conductance as a function of the basis size $B$ for different frequencies
and parameter sets (for one particular disorder realization).
Clearly, the convergence of the conductance is much slower than that of
the other quantities considered so far. However, a basis size of $B=300$
is sufficient to reduce the error to a few percent in all cases.
The upper row in Fig. \ref{fig:sig_conv} also shows that the convergence
becomes slower at higher frequencies. This is to be expected since
higher excited states are necessary to calculate the conductance at
higher frequency. We have carried out similar test calculations for
other observables of interest like the single-particle density of
states and the return probability of single-particle excitations \cite{epp_d}.
Since these quantities also involve excited states their  convergence
properties are overall very similar to that of the conductance.

\begin{figure}
  \epsfxsize=\figuresize
  \centerline{\epsffile{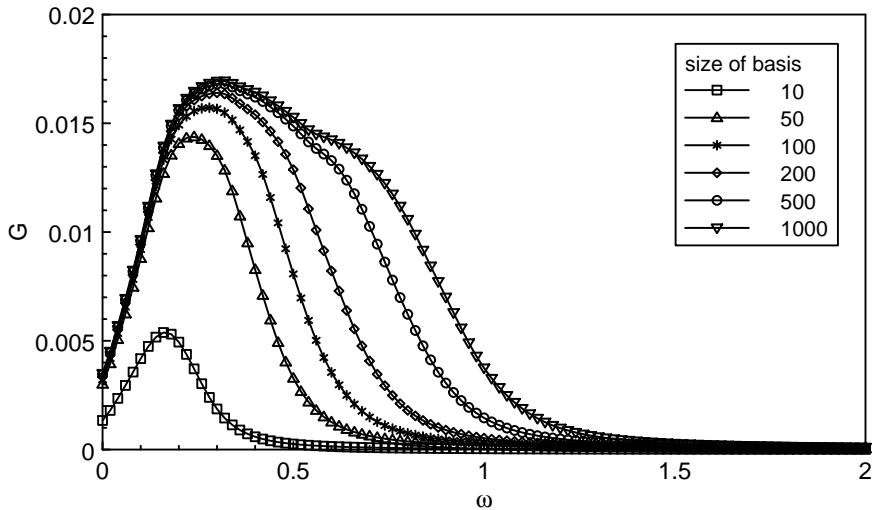}}
  \caption{Conductance as a function of frequency for a system of 18 electrons
      on  $6 \times 6$ sites, $W_0=1,~t=0.1,~U=0.1$. The data points represent
      logarithmic averages over 400 disorder realizations.
      \vspace*{1ex}}
  \label{fig:G_conv}
\end{figure}

We now turn to the question how to judge the quality of the approximation
for larger systems sizes when no exact diagonalization results are available
for comparison. In this case we use a heuristic method. We carry out several
calculations with increasing basis sizes and check how much the results
change. If this change becomes sufficiently small we stop.
In Fig.\ \ref{fig:G_conv} such a series of calculations is shown for
the conductance of a
system of 18 electrons on $6 \times 6$ lattice sites. The dimension
of the full Hilbert space is $9.1 * 10^9$.
The figure shows that with too small a basis ($B=10$) the low-frequency
conductance values are too small by a sizeable factor of around 3.
However, already a basis size of $B=100$ gives a very small error for
frequencies smaller than $\omega=0.2$. A further increase of $B$ leaves
the low-frequency part of the conductance curve essentially unchanged
but systematically improves the results at higher frequencies. This happens
because higher excited states are responsible for the conductance
at higher frequency.  With increasing basis size more and more of these
states can be calculated with sufficient accuracy.

\section{Selected results obtained with the HFD method}
\label{sec:Results}
In the last few years we have used the HFD method to study various
transport and localization properties of interacting electrons
in a random potential. The interest in this topic resurged during the
1990's because new experimental and theoretical results had cast
doubt on the established theories based on perturbation theory
for weak disorder and interactions. In this section we will
briefly summarize a few of the key results obtained.

Since a realistic description has to include the spin degrees of freedom
we have generalized the quantum Coulomb glass (\ref{eq:Hamiltonian})
to electrons with spin. A system with $\mathcal{N}$ lattice sites now
contains $N=N_\uparrow +N_\downarrow=2K \mathcal{N}$
electrons ($0\!<\!K\!<\!1$) and has a compensating positive charge of
$2Ke$ on each lattice site. The Hamiltonian is given by
\begin{eqnarray}
H &=&  -t  \sum_{\langle ij\rangle, \sigma} (c_{i\sigma}^\dagger c_{j\sigma}
      + c_{j\sigma}^\dagger c_{i\sigma})+
       \sum_{i,\sigma} \varphi_i  n_{i\sigma} \\
&&+~\frac{1}{2}\sum_{i\not=j,
       \sigma,\sigma'}U_{ij}~(n_{i\sigma}-K)(n_{j\sigma'}-K)\nonumber
     + U_{\rm H} \sum_{i} n_{i\uparrow} n_{i\downarrow}\nonumber
\label{eq:Spin_QCG}
\end{eqnarray}
where $c_{i\sigma}^\dagger$, $c_{i\sigma}$, and $n_{i\sigma}$ are the creation,
annihilation and occupation number operators for electrons at site $i$
with spin $\sigma$.
Two electrons on the same site interact via a Hubbard interaction $U_{\rm H}$.

\begin{figure}
  \epsfxsize=10cm
  \centerline{\epsffile{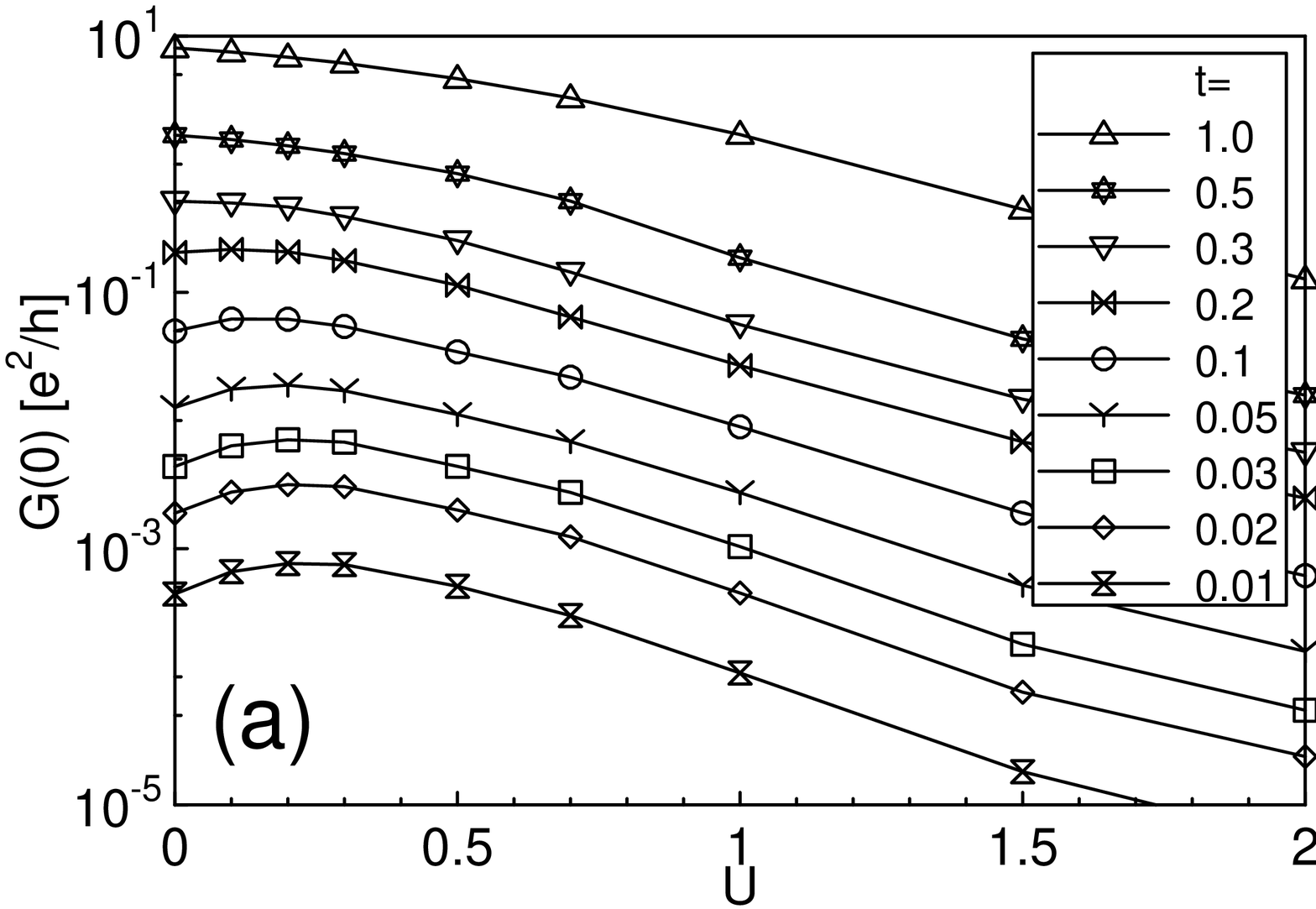}}
  \epsfxsize=10cm
  \centerline{\epsffile{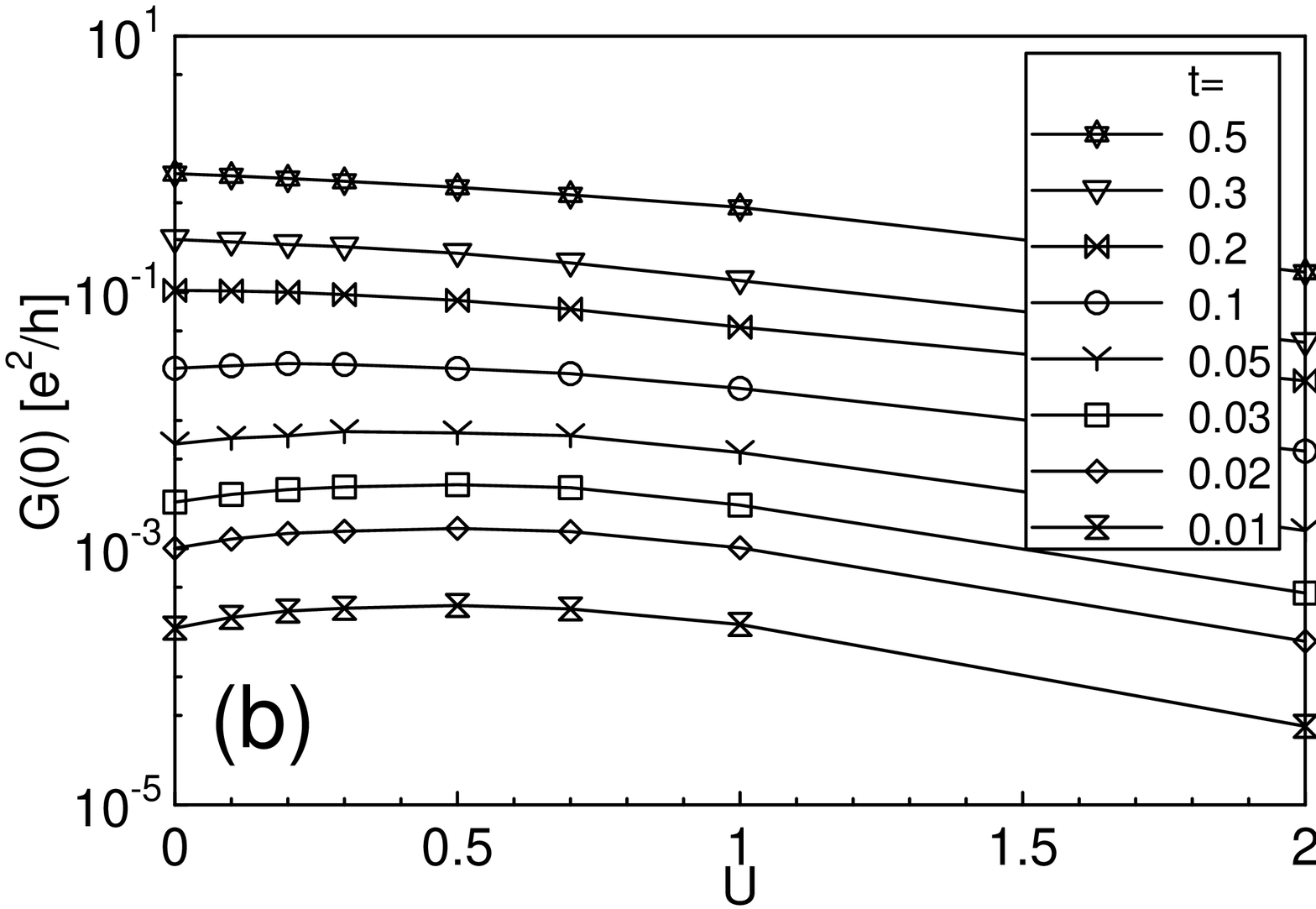}}
  \caption{d.c. conductance $G(0)$ as a function of interaction $U$
      for a system of $4 \times 4$ sites occupied by
      (a) 8 spin-up and 8 spin-down electrons and (b) 8 spinless fermions.
      The Hubbard energy is $U_H=U$, and the HFD basis size is $B=500$.
      The inhomogeneous broadening
      is $\gamma=0.0625$.}
  \label{Fig:conductance}
\end{figure}

Figure \ref{Fig:conductance} shows the typical conductance values
of a system of $4 \times 4$ lattice sites at half filling as
a function of the interaction $U$ for different hopping matrix elements $t$.
Panel (a) shows results for an occupation of 8 spin-up and 8 spin-down
electrons while panel (b) is for 8 spinless fermions.
Since the logarithm of the conductance rather than the
conductance itself is a self-averaging quantity in a disordered system,  we calculate
the typical conductance by averaging the logarithms of the conductances of 1000
(400 in the spinless case) disorder configurations.
The data show that weak electron-electron interactions
reduce the d.c. conductance for large kinetic energy $t$.
For small $t$, i.e., in the localized regime,
small and moderate interactions significantly enhance the d.c. conductance.
For larger interaction strength the conductance drops no matter what $t$ is,
indicating the crossover to a Wigner crystal or Wigner glass.
The data also show that the spin degrees of freedom do not change the
qualitative behavior in the parameter region investigated.
In fact, after rescaling the conductance of the system with spins by 1/2
and the interaction strength by 2 the two sets of curves nicely fall on top
of each other within the statistical accuracy. Therefore we conclude that
for the systems considered the Coulomb interaction
plays the essential role for the delocalizing tendency for weak
interactions as well as for the localizing tendency for strong interactions.

\begin{figure}
  \epsfxsize=10cm
  \centerline{\epsffile{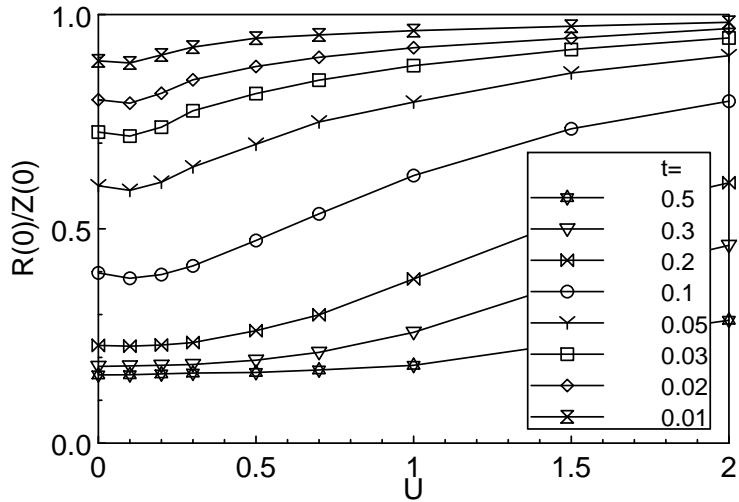}}
  \caption{Normalized single-particle return probability at the Fermi energy,
  $R_\sigma(0)/Z_\sigma(0)$ for a system of $4 \times 4$ sites occupied by
  8 spin-up and 8 spin-down electrons. The parameters are
  as in Fig. \protect\ref{Fig:conductance}}
  \label{Fig:ret_norm}
\end{figure}

In addition to the conductance we have also studied the localization
properties of single-particle excitations which we characterize
by the single-particle return probability \cite{EC1970}
\begin{equation}
R_\sigma(\omega) =  \frac 1 {g(\omega)} \frac 1 {\mathcal{N}} \sum_{i} \lim_{\delta \to 0} \frac \delta \pi \, G_{i\sigma i\sigma}^R(\omega + i \delta)
     \, G_{i\sigma i\sigma}^A(\omega - i \delta).
\label{eq:return}
\end{equation}
Here $G_{i\sigma j\sigma'}^{R,A}(\omega)$ are the retarded and advanced
single-particle Greens functions. $g(\omega)$ denotes the single-particle density of states.
In an interacting system the return
probability (\ref{eq:return})
entangles localization and decay information, since $R_\sigma$ is decreased
from 1 not only by delocalization but also by decay of the
quasi-particles. In order to extract the localization of the quasi-particles
we normalize $R_\sigma(\omega)$ by the square of the quasiparticle weight,
$Z_\sigma(\omega)$. The resulting normalized return probability is presented
in Fig.\ \ref{Fig:ret_norm}.
The figure shows that, in general, interactions
tend to localize the {\em single-particle} excitations. The delocalization
at low interaction strength and small kinetic energy is a much smaller effect
than the corresponding increase in the conductance discussed above.
This is in agreement with earlier results \cite{epper_hf}
based on
the Hartree-Fock approximation. The reason for the strong single-particle
localization is the Coulomb gap in the {\em single-particle} density of
states which effectively reduces the overlap between excitations close to the
Fermi level. The density of states of the particle-hole excitations
responsible for the conductance has a much weaker gap, if any.

For a more detailed presentation of our results
see, e.g., Refs.\ \cite{letter,giessen,pils}
for studies of spinless fermions in one, two and three dimensions
or Ref. \cite{pollak} for investigations of electrons with spin.

\section{Summary and conclusions}
\label{sec:CON}
In this paper we have discussed an efficient numerical method, the
Hartree-Fock based diagonalization, for the
simulation of interacting electrons.
It is based on diagonalizing the Hamiltonian in a subspace of the
Hilbert space spanned by the low-energy Hartree-Fock eigenstates.
The method is very general, we have
used it to study a variety of systems including quantum nano-structures
\cite{dots}. However, it is particularly suited for the investigation
of strongly disordered systems because the disorder lifts most of the
degeneracies which complicate the application of the Hartree-Fock
approximation and thus also the Hartree-Fock based diagonalization.

We have presented an evaluation of the method for the example of the
quantum Coulomb glass, a lattice model of interacting electrons in
a random potential. We have shown that the ground state occupation numbers
and the ground state energy converge very rapidly with increasing
size of the subspace used. The convergence of the ground state
energy is approximately exponential in the parameter range studied.
The convergence of dynamic observables whose calculation involves
excited states is less quick. However, we were able to achieve
satisfactory accuracy with very small subspace sizes.

In the last few years we have used the Hartree-Fock based
diagonalization to investigate a variety of transport and localization
properties of strongly disordered interacting electrons. We have
calculated the conductance, the charge stiffness and various localization
properties of single-particle excitations. We found that weak
electron-electron interactions tend to increase the transport in the
localized regime but decrease the transport in the metallic regime.
Sufficiently strong interactions always localize the system which
then forms a Wigner crystal or a Wigner glass.

This work was supported in part by the German Research Foundation under
grant no. SFB 393/C2.

\end{document}